\documentclass[conference]{IEEEtran}

\usepackage[british]{babel}
\usepackage{cite}
\usepackage{amsmath,amssymb,amsfonts,amsthm}
\usepackage{algorithmic}
\usepackage{physics}
 \usepackage{graphicx}
\usepackage{textcomp}
\usepackage[T1]{fontenc}
\usepackage{caption}
\usepackage{cases}
\usepackage{subcaption}
\usepackage{epstopdf}
\usepackage{overpic}
\usepackage{array}
\usepackage{color}
\usepackage{url}
\usepackage{bbm}
\usepackage{enumerate}
\usepackage[shortlabels]{enumitem}

\usepackage[none]{hyphenat}
\usepackage{hyperref}
\usepackage{lipsum}

\def\BibTeX{{\rm B\kern-.05em{\sc i\kern-.025em b}\kern-.08em
    T\kern-.1667em\lower.7ex\hbox{E}\kern-.125emX}}
    
\usepackage{color}
\usepackage[dvipsnames,svgnames]{xcolor}
\usepackage[textwidth=30mm]{todonotes}

\def\Complex{\mathbb{C}}

\def\imagunit{\mathsf{j}} 

\newcommand{\vect}[1]{{\boldsymbol{#1}}}

\theoremstyle{plain}

\graphicspath{{./images/}}

\begin{document}

\title{Line-of-Sight MIMO via \\ Reflection From a Smooth Surface\vspace{-0cm}}

\author{
\IEEEauthorblockN{Andrea Pizzo\IEEEauthorrefmark{1}, Angel Lozano\IEEEauthorrefmark{1}, Sundeep Rangan\IEEEauthorrefmark{2}, Thomas Marzetta\IEEEauthorrefmark{2} }
\IEEEauthorblockA{\IEEEauthorrefmark{1}\small{Department of Information and Communication Technologies, Universitat Pompeu Fabra, Spain (\{andrea.pizzo,angel.lozano\}@upf.edu)}}
\IEEEauthorblockA{\IEEEauthorrefmark{2}\small{Department of Electrical and Computer Engineering, New York University, USA (\{srangan,tom.marzetta\}@nyu.edu)}\vspace{-0cm}}
}

\maketitle

\begin{abstract}
We provide a 
deterministic channel model for a scenario where wireless connectivity is established through a reflection from a planar smooth surface of an infinite extent. The developed model is rigorously built upon the physics of wave propagation, and is as precise as tight are the unboundedness and smoothness assumptions on the surface. This model allows establishing that line-of-sight spatial multiplexing can take place via
reflection off an electrically large surface, a situation of high interest for mmWave and terahertz frequencies.

\vspace{-0cm}
\end{abstract}


\IEEEpeerreviewmaketitle

\section{Introduction}

The wealth of unexplored spectrum in the
mmWave and terahertz 
 ranges brings an onrush of wireless research seeking its fortune at these frequencies \cite{Sundeep2014,Rappaport2015,Rappaport2019}.
The short transmission range for which these frequencies are most suitable, in conjunction with the short wavelength, enable reasonably sized
arrays to access multiple spatial degrees of freedom (DOF) even under direct line-of-sight (LOS) propagation \cite{Rodwell2011}.
This potential has unleashed much research activity on LOS multiple-input multiple-output (MIMO) communication \cite{9422343}.

A downside of these high frequencies is blockage and lack of diffraction around obstacles, which may render LOS MIMO vulnerable to interruptions.
This naturally raises the interest in studying whether LOS MIMO links could also operate through a reflection, capitalizing on the availability in many environments of interest of surfaces that are electrically (i.e., relative to the wavelength) large.

This paper examines MIMO communication via reflection off a flat smooth surface of infinite extent. One possibility would be to apply ray tracing tools \cite{Xia2021}, but
the 
accuracy to which the environment should be characterized to prevent artifacts is not known a priori.
Instead, we derive a physics-based scalar channel model that is valid for arbitrary materials,
with a perfectly conductive material as a special case. 
Generalization to vector electromagnetic channels would allow incorporating the role of polarization \cite{tulino2003capacity}. 


\section{Plane-Wave Reflection}

\begin{figure}[t!]
        \centering
	\begin{overpic}[width=.85\columnwidth,tics=10]{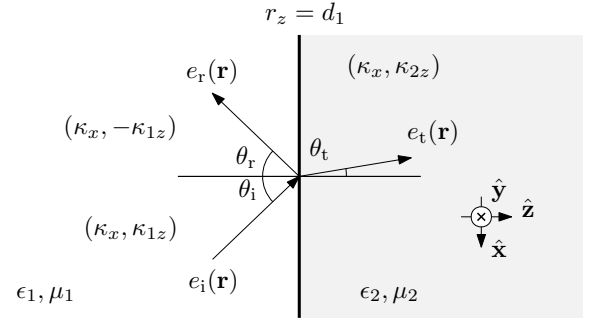}
\end{overpic} \vspace{-0.0cm}
                \caption{3D medium with a planar surface at $r_z=d_1$. View from the $xz$-plane of incidence.} \vspace{0.0cm}
                \label{fig:propagation} 
        \end{figure}  
      
We study wave propagation in a 3D isotropic and unbounded medium containing a
planar surface at $r_z=d_1$, orthogonal to an arbitrary $z$-axis. Thus, the propagation medium is divided into a left half-space $r_z<d_1$ and a right half-space $r_z>d_1$, as illustrated in Fig.~\ref{fig:propagation}. All quantities associated with the left and right half-spaces are subscripted by $1$ and $2$, respectively. We assume free-space in the left half-space, whereas the right half-space consists of a lossless material having permeability $\mu_2$ and permittivity $\epsilon_2$.

Consider the $xz$-plane 
containing the direction of propagation and the surface normal, namely the \emph{plane of incidence}. 
An upgoing incident plane wave traveling in the half-space $r_z<d_1$ from an angle $\theta_{\text i}\in [0,\pi/2]$ with respect to the surface normal is expressed as\footnote{We use $+$ and $-$ to distinguish between quantities associated with upgoing and downgoing waves, respectively. Upgoing waves $e^{\imagunit  \kappa_{1z} r_z}$ propagate along the $z$-axis whereas downgoing waves $e^{-\imagunit \kappa_{1z} r_z}$ propagate along the $- z$-axis.}
\begin{equation} \label{incident_plane_wave}
e_{\text i}(\vect{r}) =  E_{\text i}^+(\kappa_x) \,  e^{\imagunit \left( \kappa_x r_x +  \kappa_{1z} r_z \right)}
\end{equation}
with amplitude $E_{\text i}^+(\kappa_x)$ and real-valued $\kappa_{1z} = ({\kappa_1^2 - \kappa_x^2})^{1/2}$ for $|\kappa_x|\le \kappa_1$, given $\kappa_1 = \omega/c$ as the wavenumber in the left half-space with $\omega$ the frequency. 
Evanescent waves are disregarded unless otherwise stated, and $(\kappa_x,\kappa_{1z}) = (\kappa_1 \sin\theta_{\text i}, \kappa_1 \cos\theta_{\text i})$ specifies the direction of the incident wave on the plane of incidence.
After impinging on the surface, a downgoing reflected plane wave is generated in the half-space $r_z<d_1$ as
\begin{align} \label{reflected_field}
e_{\text r}(\vect{r}) &=  E_{\text r}^-(\kappa_{x,{\text r}}) \, e^{\imagunit \left(\kappa_{x,{\text r}}  r_x - \kappa_{z,{\text r}}  (r_z-2d_1)\right)}
\end{align}
with amplitude $E_{\text r}^+(\kappa_{x,{\text r}})$ and $\kappa_{z,{\text r}} = ({\kappa_1^2 - \kappa_{x,{\text r}}^2})^{1/2}$ in the domain $|\kappa_{x,{\text r}}|\le \kappa_1$. Here, $(\kappa_{x,{\text r}},\kappa_{z,{\text r}}) = (\kappa_1 \sin\theta_{\text r},\kappa_1 \cos\theta_{\text r})$ where $\theta_{\text r}$ is the reflection angle relative to the surface normal. With respect to \eqref{incident_plane_wave}, the extra phase in \eqref{reflected_field} accounts for the round-trip delay accumulated by the incident plane wave during its travel to the surface and back, along the $z$-axis. 
This can be regarded as a migration of the reflected plane wave and is directly connected to the image theorem, as discussed in Sec.~\ref{sec:image_th}.
In addition to \eqref{reflected_field}, the upgoing plane wave in \eqref{incident_plane_wave} generates in the half-space $r_z>d_1$  a transmitted wave
\begin{align} \label{transmitted_field}
e_{\text t}(\vect{r}) &=   E_{\text t}^+(\kappa_{x,{\text t}}) \, e^{\imagunit \left(\kappa_{x,{\text t}} r_x  + \kappa_{z,{\text t}} (r_z - d_1)\right)} e^{\imagunit \kappa_{1z} d_1}
\end{align}
with amplitude $E_{\text t}^+(\kappa_{x,{\text t}})$ and $\kappa_{z,{\text t}} = ({\kappa_2^2 - \kappa_{x,{\text t}}^2})^{1/2}$, given $\kappa_2 = \omega/c_2$ with $c_2 = c/n_2$ the wave speed in the right half-space and $n_2 > 1$ the corresponding refractive index. Here, $(\kappa_{x,{\text t}},\kappa_{z,{\text t}}) = (\kappa_2 \sin\theta_{\text t},\kappa_2 \cos\theta_{\text t})$ with $\theta_{\text t}$ the transmitted angle with respect to the surface normal in the right half-space.
Similarly to \eqref{reflected_field}, we added the migration effect due to the one-way delay accumulated by the incident wave in its travel from the origin to the surface, along the $z$-axis. 
A complete description of 
what unfolds 
on the incident plane is obtained by combining \eqref{incident_plane_wave} and \eqref{reflected_field} into
\begin{equation} \label{sum_field_tot}
e(\vect{r}) = 
\begin{cases}
e_{\text i}(\vect{r}) + e_{\text r}(\vect{r}) & \quad r_z < d_1 \\
e_{\text t}(\vect{r})  & \quad r_z > d_1.
\end{cases}
\end{equation}
The boundary conditions 
require continuity of the field and its first-order derivative across the surface \cite[Eq.~2.1.9a]{ChewBook}, namely
\begin{numcases}{}  
e(\vect{r})|_{r_z = d_1^-} = e(\vect{r})|_{r_z = d_1^+} \label{boundary_condition_1} \\ 
\mu_1^{-1} \frac{\partial}{\partial r_z} e(\vect{r})|_{r_z = d_1^-} = \mu_2^{-1} \frac{\partial}{\partial r_z} e(\vect{r})|_{r_z = d_1^+} \label{boundary_condition_2}
\end{numcases}
for all $(r_x,r_y)$. Here, \eqref{boundary_condition_1} leads to the so-called phase matching condition
\begin{equation} \label{phase_matching}
 \kappa_{x,{\text r}} = \kappa_{x,{\text t}} = \kappa_x
\end{equation}
from which $\kappa_{z,{\text r}} = \kappa_{1z}$ and $\kappa_{z,{\text t}} = ({\kappa_2^2 - \kappa_x^2})^{1/2} = \kappa_{2z}$. Notice that, since $n_2 > n_1$ (i.e., $\kappa_2 > \kappa_1$), we have that $\kappa_{2z}$ is real-valued for $|\kappa_x|\le \kappa_1$. Hence, no evanescent waves are created in the right half-space;
the transmitted wave is still propagating.
In spherical coordinates, we would obtain the general Snell's law of refraction \cite[Eq.~1.5.6]{OpticsBook}, $\theta_{\text r} = \theta_{\text i}$ and $\frac{\sin(\theta_{\text t})}{\sin(\theta_{\text i})} = \frac{n_1}{n_2}$, for all incident angles.
Also, plugging \eqref{phase_matching} into \eqref{boundary_condition_1} and accounting for  \eqref{boundary_condition_2} we obtain, for all $\kappa_x $,
\begin{equation}  \label{system}
\begin{cases}
E_{\text i}^+(\kappa_x) + E_{\text r}^-(\kappa_x) = E_{\text t}^+(\kappa_x) \\
E_{\text i}^+(\kappa_x) - E_{\text r}^-(\kappa_x) = \alpha(\kappa_x) E_{\text t}^+(\kappa_x) 
\end{cases}
\end{equation} 
where we defined $\alpha(\kappa_x)  = {\mu_1 \kappa_{2z}}/({\mu_2 \kappa_{1z}})$.
Hence, the boundary conditions form a system of two equations in two unknowns. 
The unknowns $E_{\text r}^-(\kappa_x)$ and $E_{\text t}^+(\kappa_x)$ are parametrized by $E_{\text i}^+(\kappa_x)$ for all $\kappa_x$. After normalization, these can be written in terms of the Fresnel reflection and transmission coefficients
\begin{equation}  \label{Fresnel_coefficients}
R_-(\kappa_x)  = \frac{E_{\text r}^-(\kappa_x)}{E_{\text i}^+(\kappa_x)} \qquad\quad 
T_+(\kappa_x)  = \frac{E_{\text t}^+(\kappa_x)}{E_{\text i}^+(\kappa_x)} ,
\end{equation}
which specify the fraction of incident field that is reflected from or transmitted across the surface, for all possible incident angles. Since the material cannot amplify the signal, the magnitudes of the Fresnel coefficients are always less than one. 
General expressions for these coefficients may be found by solving \eqref{system}, e.g., using Cramer's rule, as \cite[Eq.~2.1.13]{ChewBook}
\begin{align} \label{reflection_coeff} 
R_-(\kappa_x) &= \frac{1-\alpha}{1+\alpha} = \frac{\mu_2 \kappa_{1z} - \mu_1 \kappa_{2z}}{\mu_2 \kappa_{1z} + \mu_1 \kappa_{2z}} \\ \label{transmission_coeff}
T_+(\kappa_x) &= \frac{2}{1+\alpha} = \frac{2 \mu_2 \kappa_{1z}}{\mu_2 \kappa_{1z} + \mu_1 \kappa_{2z}}.
\end{align}
Both relate through the linear relationship
\begin{equation} \label{reflection_transmission}
1 + R_-(\kappa_x) = T_+(\kappa_x).
\end{equation}
In a homogeneous medium, $\mu_1=\mu_2$ and $\epsilon_1=\epsilon_2$ (i.e., $\kappa_{1z}=\kappa_{2z}$) so that $R_-(\kappa_x)=0$ and $T_+(\kappa_x)=1$. With a perfectly conductive material, $\mu_2 = 0$ so that $R_-(\kappa_x)=-1$ and $T_+(\kappa_x)=0$, corresponding to total reflection.

 \begin{figure*}[t!]
\begin{equation} \tag{16} \label{received_field_total}
e(\vect{r}) = 
\begin{cases} \displaystyle
 \iint_{{\mathcal{D}}}    \left(E_{\text i}^-(\kappa_x,\kappa_y) + E_{\text i}^+(\kappa_x,\kappa_y) R_-(\kappa_x,\kappa_y) \, e^{- \imagunit \kappa_{1z} (r_z-2d_1)}\right)  e^{\imagunit \left( \kappa_x r_x + \kappa_y r_y\right)} \frac{d\kappa_x}{2\pi}\frac{d\kappa_y}{2\pi} \quad r_z < -R_0 \\ \displaystyle
 \iint_{{\mathcal{D}}}  E_{\text i}^+(\kappa_x,\kappa_y) \left(e^{\imagunit \kappa_{1z} r_z} +  R_-(\kappa_x,\kappa_y) e^{- \imagunit \kappa_{1z} (r_z-2d_1)}\right)  e^{\imagunit \left( \kappa_x r_x + \kappa_y r_y\right)} \frac{d\kappa_x}{2\pi}\frac{d\kappa_y}{2\pi}  \quad\quad\;\; R_0 < r_z \le d_1 \\ \displaystyle
  \iint_{{\mathcal{D}}} E_{\text i}^+(\kappa_x,\kappa_y) e^{\imagunit \kappa_{1z} d_1} T_+(\kappa_x,\kappa_y)  e^{\imagunit \kappa_{2z} (r_z-d_1)} e^{\imagunit \left( \kappa_x r_x + \kappa_y r_y\right)} \frac{d\kappa_x}{2\pi}\frac{d\kappa_y}{2\pi}   \qquad\qquad\qquad\quad r_z>d_1
\end{cases}
\end{equation}
\hrule\vspace{-0cm}
\end{figure*}

\section{Fourier Plane-wave Representation}

Studying the interaction between a plane wave and a smooth surface is of key importance as the field generated by an arbitrary source can be represented exactly in terms of plane waves---even in the near field \cite{ChewBook,PlaneWaveBook}. Particularly, let $j(\vect{r})$ be a narrowband source 
defined within a volume $V_S \subset \{r_z<d_1\}$. This creates an incident field, $e_{\text i}(\vect{r})$, that can be expanded exactly as an integral superposition of plane waves, propagating and evanescent \cite{ChewBook,PlaneWaveBook}. If $V_S$ is embedded in a sphere of radius $R_0 <d_1$, then, outside of this sphere, on the plane of incidence \cite{PizzoIT21} 
 \begin{align}  \label{incident_field}
e_{\text i}(\vect{r}) &  =
\begin{cases} \displaystyle
\int_{-\infty}^{\infty}   E_{\text i}^-(\kappa_x)  \,  e^{\imagunit \left( \kappa_x r_x -  \kappa_{1z} r_z\right)} \, \frac{d\kappa_x}{2\pi} 
\qquad r_z < -R_0\\\displaystyle
\int_{-\infty}^{\infty}   E_{\text i}^+(\kappa_x)  \, e^{\imagunit \left( \kappa_x r_x +  \kappa_{1z} r_z \right)} \, \frac{d\kappa_x}{2\pi}
\qquad r_z >  R_0
\end{cases}
\end{align}
where each plane wave has complex-valued amplitude
\begin{equation} \label{incident_spectrum}
E_{\text i}^\pm(\kappa_x) =  \frac{\kappa_1 \eta_1}{2} \frac{J_\pm(\kappa_x)}{\kappa_{1z}}  
\end{equation}
with $J_\pm(\kappa_x)$ the wavenumber spectrum of $j(\vect{r})$ obtained via a two-dimensional Fourier transform on the $xz$-plane evaluated at $\kappa_z =  \pm \kappa_{1z}$, i.e.,
\begin{equation} \label{source_spectrum}
J_\pm(\kappa_x) = \iint_{V_S}    j(\vect{s}) \, e^{-\imagunit \left(\kappa_x s_x \pm \kappa_{1z} s_z\right)} \, ds_x ds_z
\end{equation}
given $\eta_1 = \sqrt{\mu_1/\epsilon_1}$. 

\section{Channel Impulse Response} 

\begin{figure*}[t!]
\begin{equation} \tag{20} \label{wavenumber_response}
H(\kappa_x,\kappa_y) = 
\begin{cases} \displaystyle
\frac{\kappa_1 \eta_1}{2}  \frac{\mathbbm{1}_{\mathcal{D}(\kappa_x,\kappa_y)}}{\kappa_{1z}}  \left(e^{-\imagunit \kappa_{1z} (r_z-s_z)}+ R_-(\kappa_x,\kappa_y) e^{-\imagunit \kappa_{1z} (r_z+s_z - 2d_1)} \right) & \quad r_z < -R_0 \\ \displaystyle
\frac{\kappa_1 \eta_1}{2}  \frac{\mathbbm{1}_{\mathcal{D}(\kappa_x,\kappa_y)}}{\kappa_{1z}}   \left(e^{\imagunit \kappa_{1z} (r_z-s_z)}  +  R_-(\kappa_x,\kappa_y)  e^{-\imagunit  \kappa_{1z} (r_z+s_z- 2d_1)} \right) & \quad\!\! R_0 < r_z \le d_1 \\ \displaystyle
\frac{\kappa_1 \eta_1}{2}  \frac{\mathbbm{1}_{\mathcal{D}(\kappa_x,\kappa_y)}}{\kappa_{1z}} e^{\imagunit \kappa_{1z} (d_1 - s_z)} T_+(\kappa_x,\kappa_y)  e^{\imagunit \kappa_{2z} (r_z - d_1)}    & \quad r_z > d_1
\end{cases}
\end{equation}
\hrule\vspace{-0.5cm}
\end{figure*}

Fundamental principles describing the reflection and transmission phenomena at the surface can be applied to each incident plane wave separately and then combined to obtain the general field expression $e(\vect{r})$.
Provided the source and the surface are separated by several wavelengths, only propagating plane waves contribute.
The wavenumber domain is limited to a compact support $(\kappa_x,\kappa_y)\in\mathcal{D}$ with $\mathcal{D}$ a disk of radius $\kappa_1$. The Fourier plane-wave representation of $e(\vect{r})$ is given in \eqref{received_field_total} with real-valued
\setcounter{equation}{16} 
\begin{equation} \label{kappaz_i}
\kappa_{iz} = \sqrt{\kappa_i^2 - \kappa_x^2 - \kappa_y^2}
\end{equation}
for $i=1,2$ while $R_-(\kappa_x,\kappa_y)$ and $T_+(\kappa_x,\kappa_y)$ are the Fresnel coefficients in \eqref{reflection_coeff} and \eqref{transmission_coeff}. Field continuity at $r_z=d_1$ can be verified in \eqref{received_field_total} by invoking \eqref{reflection_transmission}. The 
relationship between $e(\vect{r})$ and  $j(\vect{s})$ is the spatial convolution
\begin{equation} \label{convolution}
e(\vect{r}) = \int_{V_S}    j(\vect{s}) \, h(\vect{r},\vect{s}) \, d\vect{s}
\end{equation}  
where $h(\vect{r},\vect{s})$ is the channel impulse response. 
Plugging \eqref{incident_spectrum} into \eqref{received_field_total} and replacing the source spectrum with its Fourier transform in \eqref{source_spectrum}, the channel response can be written as the inverse Fourier transform
\begin{equation} \label{impulse_response_Fourier}
h(x,y;r_z,s_z) = \iint_{-\infty}^{\infty}  \!\!\! H(\kappa_x,\kappa_y)  e^{\imagunit \left(\kappa_x x +  \kappa_y y \right)} \, \frac{d\kappa_x}{2\pi} \frac{d\kappa_y}{2\pi} 
\end{equation}
of the spectrum in \eqref{wavenumber_response}, where we have embedded the integration domain 
into a functional dependence through an indicator function.
This response is a function of the space-lag variables $r_x-s_x = x$ and $r_y-s_y = y$.

\section{Image Theorem} \label{sec:image_th}

Recalling \eqref{incident_spectrum}, we can rewrite the reflected component in \eqref{received_field_total} as
\setcounter{equation}{20} 
 \begin{align}   \notag
e_{\text r}(\vect{r}) =  & \frac{\kappa_1 \eta_1}{2} \iint_{-\infty}^{\infty}    \frac{J_{\text r}^-(\kappa_x,\kappa_y)}{\kappa_{1z}} \, e^{\imagunit \kappa_{1z} 2d_1}  \\ & \hspace{0cm} \label{reflected_field_image}
\cdot  e^{\imagunit \left( \kappa_x r_x + \kappa_y r_y -  \kappa_{1z} r_z\right)} \frac{d\kappa_x}{2\pi}\frac{d\kappa_y}{2\pi}
\qquad\quad r_z \le d_1
\end{align}
where $J_{\text r}^-(\kappa_x,\kappa_y)$ relates to the spectrum $J_+(\kappa_x,\kappa_y)$ of the original source in \eqref{source_spectrum} via
\begin{equation} \label{equivalent_source_reflection_pec}
J_{\text r}^-(\kappa_x,\kappa_y) = J_+(\kappa_x,\kappa_y) R_-(\kappa_x,\kappa_y).
\end{equation}
Notice that \eqref{reflected_field_image} and the 3D counterpart of \eqref{incident_field} have the same form, up to a phase shift. 
Here, $J_{\text r}^-(\kappa_x,\kappa_y)$ is the spectrum in \eqref{source_spectrum} of a fictitious source $j_{\text r}^-(\vect{r})$. 
The phase shift in \eqref{reflected_field_image} is equivalently a shift in the centroid of the source around $r_z=2d_1$. 
For a perfectly conductive surface, i.e., such that $R_-(\kappa_x,\kappa_y)=-1$, the reflected field in \eqref{reflected_field_image} may be reproduced exactly by replicating the original source at $r_z = 2d_1$. This is the \emph{image theorem} whereby the ideal reflection elicited by a perfect conductor is equivalent to a mirror
image of the original source 
 \cite[Sec.~4.7.1]{BalanisBook}.
As an example, for a point source $j(\vect{r}) = \delta(\vect{r})$, applying Weyl's identity \cite[Eq.~2.2.27]{ChewBook}
\begin{equation} \label{Weyl_identity}
\frac{e^{\imagunit \kappa_1 \sqrt{r_x^2 + r_y^2 + r_z^2}}}{\sqrt{r_x^2 + r_y^2 + r_z^2}}= \frac{\imagunit}{2\pi}  \iint_{-\infty}^{\infty} \frac{e^{\imagunit ( \kappa_x x + \kappa_y y + \kappa_{1z} |z|)}}{\kappa_{1z}}  \, d\kappa_xd\kappa_y
\end{equation}
we obtain
\begin{equation} \label{field_point_source}
 e(\vect{r}) \propto \frac{e^{\imagunit \kappa_1 \sqrt{r_x^2 + r_y^2 + r_z^2}}}{\sqrt{r_x^2 + r_y^2 + r_z^2}} +
 \frac{e^{\imagunit \kappa_1 \sqrt{r_x^2 + r_y^2 + (r_z-2 d_1)^2}}}{\sqrt{r_x^2 + r_y^2 + (r_z-2 d_1)^2}} ,
 \end{equation}
which describes two spherical waves generated by the original point source and its image. 
For an arbitrary material, the equivalent source is the inverse Fourier transform of \eqref{equivalent_source_reflection_pec}.
This simplifies when the surface is far enough from the source that $R_-(\kappa_x,\kappa_y)$ in \eqref{equivalent_source_reflection_pec} is roughly constant across the receive array and for all possible incident angles; then, the equivalent source becomes a weakened (and phase-shifted) version of the original one, which is the premise of ray tracing algorithms.
However, this need not be the case in LOS MIMO, which rests on the range being short.


\section{Application to MIMO Communication}

\begin{figure}
        \centering
	\begin{overpic}[width=.999\columnwidth,tics=10]{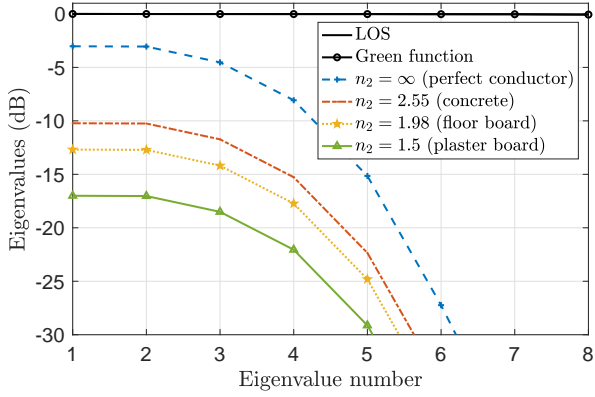}
\end{overpic} \vspace{-0.3cm}
                \caption{Normalized channel eigenvalues 
                for materials chosen from Table~\ref{tab:refractive_indices}. Parallel ULAs separated by $D = 10$~m and equipped with $N=8$ antennas, with antenna spacing $d(D)$ in \eqref{Rayleigh}. The surface is at $d_1 = 15$~m from the transmitter.}
                \vspace{0.0cm}
                \label{fig:eigenvalues_n2_LoS} 
        \end{figure}   
        
        \begin{figure}
\centering\vspace{0.0cm}
\includegraphics[width=.999\linewidth]{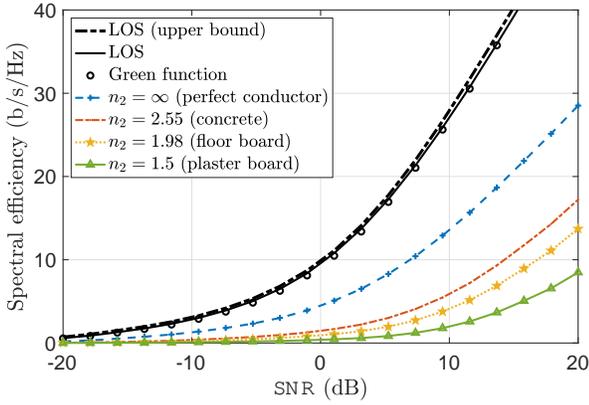} \vspace{-0.3cm}
\caption{Spectral efficiency 
as a function of SNR for different materials.
The antenna spacing is $d(D,{\sf SNR})$ in \eqref{spacing}.
}\vspace{0.0cm}
\label{fig:capacity_n2_LoS}
\end{figure}
        
We now apply the developed channel model to numerically evaluate the channel eigenvalues, DOF, and spectral efficiency. 
With $N$ transmit and $N$ receive antennas, the channel matrix $\vect{H} \in\Complex^{N \times N}$ is obtained by sampling the response at the antenna locations, 
$[\vect{H}]_{m,n} = h(\vect{r}_m,\vect{s}_n)$
for $m,n=1, \dots, N$. We consider uniform linear arrays (ULAs) at $57.5$~GHz under the proviso of these ULAs being substantially shorter than their separating range, the so-called paraxial approximation, so we can leverage results available for LOS channels \cite{HeedongISIT,Heedong2021}. We will show how these generalize to reflected transmissions.
We hasten to emphasize that the reliance on the paraxial approximation is confined to the production of benchmark results for LOS MIMO;
our channel model is valid regardless of the geometry.
The frequency, in turn, is motivated by mmWave applications \cite{Sundeep2014} and by the availability of refractive indexes for most common materials (see Table~\ref{tab:refractive_indices}) \cite{Sato}.

\begin{table}
        \caption{Refractive indices of common building materials at $57.5$~GHz \cite{Sato}.\vspace{-0cm}}  \label{tab:refractive_indices}
\centering
    \begin{tabular}{|c|c|c|} 
    \hline
    {\bf Material} & {\bf $n_2$} & {\bf $\kappa_2$ [Krad/m]}    \\      \hline\hline 
    Perfect conductor & $\infty$ & $\infty$ \\ \hline
    Concrete &  $2.55$ &  $3.07$   \\  \hline
    Floor board & $1.98$  &  $2.38$  \\  \hline
    Plaster board & $1.50$ &  $1.81$  \\ \hline  
    \end{tabular}
\end{table}

The ULAs are aligned with the $x$-axis and equipped with $N=8$ antennas with spacing $d$. 
The range between the arrays is $D = 10$~m whereas the surface is at $d_1= 15$~m.
We begin by validating the developed model for the direct LOS channel, as there is an explicit solution in this case, namely the Green's function. 
We set
\begin{equation} \label{Rayleigh}
d(D)=\sqrt{\lambda D/N},
\end{equation}
which renders the direct $\vect{H}$ a Fourier matrix and is optimum at high SNR 
\cite{Rodwell2011,HeedongISIT}. The normalized channel eigenvalues of $\vect{H} \vect{H}^*$, $\lambda_n(\vect{H})$, are plotted in Fig.~\ref{fig:eigenvalues_n2_LoS}. 
The perfect match validates the developed model for this LOS setting.

For the second term in \eqref{wavenumber_response}, which corresponds to the reflection, the eigenvalues are also shown in Fig.~\ref{fig:eigenvalues_n2_LoS}.
 These undergo two effects relative to their LOS brethren:
 \begin{itemize}
 \item \emph{Power loss} caused by the longer range (higher pathloss) and by the reflection of only a share of the incident power, with denser materials reflecting better. 
 \item \emph{Spatial selectivity} due the antenna spacing in \eqref{Rayleigh} being suboptimal for the longer range of the reflected channel.
 \end{itemize}

We now gauge the capacity with channel-state information at transmitter and receiver, which equals \cite{tulino2004mimo,LozanoBook}
\begin{align} \label{capacity}
C(\vect{H},{\sf SNR}) =\sum_{n=1}^{N} \log_2 \! \left(1 +  \left(\nu - \frac{1}{\lambda_n(\vect{H})} \right)^{\!\!+} \!\!\! \lambda_n(\vect{H}) \right)
\end{align}
where $\nu$ is such that $\sum_{n=1}^{N} \left(\nu - {1}/{\lambda_n(\vect{H})} \right)^+ = {\sf SNR}$ with the eigenvalues normalized such that $\sum_{n=1}^N \lambda_n(\vect{H}) = N^2$.
At a given SNR, $\max_{\vect{H}} C(\vect{H},{\sf SNR}) \leq C({\sf SNR})$ 
with \cite{HeedongISIT,Heedong2021}
\begin{equation} \label{upper_bound}
C({\sf SNR}) = \max_{\rho \in\{1,2, \ldots, N\}} \rho\log_2\left(1 + \frac{{\sf SNR}}{\rho}  \frac{N^2}{\rho} \right).
\end{equation}
This capacity upper bound corresponds to $\rho$ nonzero identical eigenvalues 
and to the SNR-dependent antenna spacing 
\begin{equation} \label{spacing}
d(D,{\sf SNR}) = \sqrt{\eta \lambda D/N} ,
\end{equation} 
for a fraction $\eta = \rho({\sf SNR})/N \in [0,1]$ of the $N$ potential DOF. 
   
The capacity $C(\vect{H},{\sf SNR})$  is reported in Fig.~\ref{fig:capacity_n2_LoS} for the antenna spacing, $d(D,{\sf SNR})$, that is optimum for the upper bound of the LOS channel at every SNR.
With respect to the LOS case, the capacity of the reflected channel experiences an offset (the power loss, due to the higher pathloss) and a reduced slope (the DOF loss, due to the spatial selectivity).

\begin{figure}
        \centering
	\begin{overpic}[width=.999\columnwidth,tics=10]{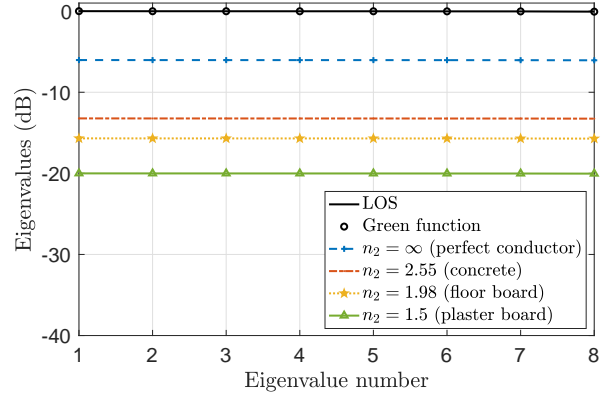}
\end{overpic} \vspace{-0.3cm}
                \caption{Normalized channel eigenvalues. 
                The antenna spacing is $d(D)$ 
                for the LOS channel and $d(D_{\text e})$ for the reflected channel.}
                \vspace{0.0cm}
                \label{fig:eigenvalues_n2_Rayleigh} 
        \end{figure}     

\begin{figure}
\centering\vspace{0.0cm}
\includegraphics[width=.999\linewidth]{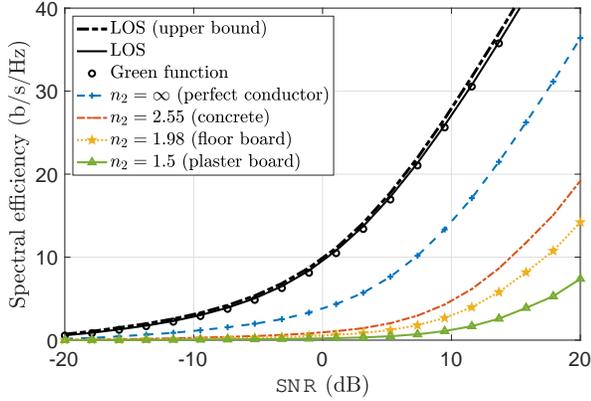} \vspace{-0.3cm}
\caption{Spectral efficiency corresponding to the eigenvalues in Fig.~\ref{fig:eigenvalues_n2_Rayleigh}. The antenna spacing is $d(D,{\sf SNR})$
for the LOS channel and $d(D_{\text e},{\sf SNR})$ for the reflected channel.
}\vspace{0.0cm}
\label{fig:capacity_n2}
\end{figure}

While the power loss is inevitable, because of the longer range, the
 spatial selectivity can be compensated by adjusting the antenna spacing. 
To this end, recall from the image theorem that the reflected channel can be regarded as an equivalent LOS channel with augmented distance $D_{\text e} > D$ between the (equivalent) source and receiver; in our setting, $D_{\text e}= 2 d_1 - D$. For the perfect conductor case, this alone justifies the choice of an antenna spacing equal to $d(D_{\text e}  )$. 
The argument is somewhat more involved for arbitrary materials, but it ultimately leads to the same result, as illustrated in Fig.~\ref{fig:eigenvalues_n2_Rayleigh}. 
Numerically, this is supported by the invariance of the curves for different materials in Fig.~\ref{fig:eigenvalues_n2_LoS}. Physically, it is explained by the paraxial approximation, based on which the field reaches the surface in a small neighborhood of a certain direction and has thus an almost constant wavenumber response. Analytically, we can pull out $R_-(\kappa_x,\kappa_y)$ in \eqref{wavenumber_response} so that all the LOS eigenvalues are identically scaled by a factor $|R_-(\kappa_x,\kappa_y)|^2$, the \emph{reflectivity} \cite[Sec.~1.5.3]{OpticsBook}.
 
 For completeness, we plot in Fig.~\ref{fig:capacity_n2} the spectral efficiency corresponding to the eigenvalues in Fig.~\ref{fig:eigenvalues_n2_Rayleigh}. With respect to Fig.~\ref{fig:capacity_n2_LoS}, the antenna spacing is set to $d(D_{\text e} ,{\sf  SNR})$ to provide full DOF via the reflection at high SNR, which means the same slope as the LOS channel in this regime.

\section{Summary}


Through a physics-based formulation, without relying on ray tracing, we have confirmed that
reflections off large and smooth flat surfaces, say walls or the ceiling, can behave as LOS links from the standpoint of MIMO communication.
Such reflections may therefore provide welcome alternative paths for LOS MIMO transmissions.
With respect to a direct LOS link, a reflected counterpart exhibits:
\begin{itemize}
\item A power loss determined by the additional range and by the share of incident power reflected by the surface.
\item A reduction in the number of DOF because of the antenna spacing tailored to the LOS link being smaller than the one that the reflected link would require at the same SNR.
\end{itemize}
If the arrays are outright configured for the reflected transmission, then the second effect is corrected with a mere adjustment of the antenna spacings for the extended range.

The above observations bode well for flexible LOS MIMO communication aided by reflections, with further work required to determine the impact of surface finiteness and roughness, and of multiple reflections.

\section*{Acknowledgment}

Work supported by the European Union-NextGenerationEU, by the European Research Council under the H2020 Framework Programme/ERC grant agreement 694974, by ICREA, and by the Fractus-UPF Chair on Tech Transfer and 6G.

\bibliographystyle{IEEEtran}
\bibliography{IEEEabrv,refs}

\begin{thebibliography}{10}
\providecommand{\url}[1]{#1}
\csname url@samestyle\endcsname
\providecommand{\newblock}{\relax}
\providecommand{\bibinfo}[2]{#2}
\providecommand{\BIBentrySTDinterwordspacing}{\spaceskip=0pt\relax}
\providecommand{\BIBentryALTinterwordstretchfactor}{4}
\providecommand{\BIBentryALTinterwordspacing}{\spaceskip=\fontdimen2\font plus
\BIBentryALTinterwordstretchfactor\fontdimen3\font minus
  \fontdimen4\font\relax}
\providecommand{\BIBforeignlanguage}[2]{{%
\expandafter\ifx\csname l@#1\endcsname\relax
\typeout{** WARNING: IEEEtran.bst: No hyphenation pattern has been}%
\typeout{** loaded for the language `#1'. Using the pattern for}%
\typeout{** the default language instead.}%
\else
\language=\csname l@#1\endcsname
\fi
#2}}
\providecommand{\BIBdecl}{\relax}
\BIBdecl

\bibitem{Sundeep2014}
S.~Rangan, T.~S. Rappaport, and E.~Erkip, ``Millimeter-wave cellular wireless
  networks: Potentials and challenges,'' \emph{Proc. IEEE}, vol. 102, no.~3,
  pp. 366--385, 2014.

\bibitem{Rappaport2015}
T.~S. Rappaport, G.~R. MacCartney, M.~K. Samimi, and S.~Sun, ``Wideband
  millimeter-wave propagation measurements and channel models for future
  wireless communication system design,'' \emph{IEEE Trans. Commun.}, vol.~63,
  no.~9, pp. 3029--3056, 2015.

\bibitem{Rappaport2019}
T.~S. Rappaport, Y.~Xing, O.~Kanhere, S.~Ju, A.~Madanayake, S.~Mandal,
  A.~Alkhateeb, and G.~C. Trichopoulos, ``Wireless communications and
  applications above 100 {GHz}: Opportunities and challenges for {6G} and
  beyond,'' \emph{IEEE Access}, vol.~7, pp. 78\,729--78\,757, 2019.

\bibitem{Rodwell2011}
E.~Torkildson, U.~Madhow, and M.~Rodwell, ``Indoor millimeter wave {MIMO}:
  Feasibility and performance,'' \emph{IEEE Trans. Wireless Commun.}, vol.~10,
  no.~12, pp. 4150--4160, 2011.

\bibitem{9422343}
H.~Do, S.~Cho, J.~Park, H.-J. Song, N.~Lee, and A.~Lozano, ``Terahertz
  line-of-sight {MIMO} communication: Theory and practical challenges,''
  \emph{IEEE Commun. Magazine}, vol.~59, no.~3, pp. 104--109, 2021.

\bibitem{Xia2021}
W.~Xia, S.~Rangan, M.~Mezzavillla, A.~Lozano, G.~Geraci, V.~Semkin, and
  G.~Loianno, ``Generative neural network channel modeling for millimeter-wave
  {UAV} communication,'' \emph{CoRR}, vol. abs/2012.09133, 2021. Online:
  \url{https://arxiv.org/abs/2012.09133}.

\bibitem{tulino2003capacity}
A.~Tulino, S.~Verdu, and A.~Lozano, ``Capacity of antenna arrays with space,
  polarization and pattern diversity,'' in \emph{IEEE Inform. Theory Workshop
  (ITW)}, 2003, pp. 324--327.

\bibitem{ChewBook}
W.~C. Chew, \emph{Waves and Fields in Inhomogenous Media}.\hskip 1em plus 0.5em
  minus 0.4em\relax Wiley-IEEE Press, 1995.

\bibitem{OpticsBook}
M.~Born and E.~Wolf, \emph{Principles of Optics}, 6th~ed.\hskip 1em plus 0.5em
  minus 0.4em\relax Pergamon Press, 1980.

\bibitem{PlaneWaveBook}
T.~B. Hansen and A.~D. Yaghjian, \emph{Plane-Wave Theory of Time-Domain
  Fields}.\hskip 1em plus 0.5em minus 0.4em\relax Wiley-IEEE Press, 1999.

\bibitem{PizzoIT21}
A.~Pizzo, L.~Sanguinetti, and T.~L. Marzetta, ``Spatial characterization of
  electromagnetic random channels,'' \emph{CoRR}, vol. abs/2103.15666, 2021.
  Online: \url{https://arxiv.org/abs/2103.15666}.

\bibitem{BalanisBook}
C.~A. Balanis, \emph{Antenna Theory: Analysis and Design}, 4th~ed.\hskip 1em
  plus 0.5em minus 0.4em\relax Wiley-Interscience, 2005.

\bibitem{HeedongISIT}
H.~Do, N.~Lee, and A.~Lozano, ``Capacity of line-of-sight {MIMO} channels,'' in
  \emph{IEEE Int. Symp. Inf. Theory (ISIT)}, 2020, pp. 2044--2048.

\bibitem{Heedong2021}
------, ``Reconfigurable {ULAs} for line-of-sight {MIMO} transmission,''
  \emph{IEEE Trans. Wireless Commun.}, vol.~20, no.~5, pp. 2933--2947, 2021.

\bibitem{Sato}
K.~Sato, H.~Kozima, H.~Masuzawa, T.~Manabe, T.~Ihara, Y.~Kasashima, and
  K.~Yamaki, ``Measurements of reflection characteristics and refractive
  indices of interior construction materials in millimeter-wave bands,'' in
  \emph{IEEE Veh. Techn. Conf.}, vol.~1, 1995, pp. 449--453 vol.1.

\bibitem{tulino2004mimo}
A.~Tulino, A.~Lozano, and S.~Verdu, ``{MIMO} capacity with channel state
  information at the transmitter,'' in \emph{IEEE Int. Symp. Spread Spectrum
  Tech. Appl. (ISSSTA)}, 2004.

\bibitem{LozanoBook}
R.~W. Heath~Jr. and A.~Lozano, \emph{Foundations of {MIMO}
  Communication}.\hskip 1em plus 0.5em minus 0.4em\relax Cambridge University
  Press, 2018.

\end{thebibliography}

\end{document}